\documentstyle[12pt]{article}

\begin{document}


\begin{titlepage}
\vskip0.5cm
\begin{center}
  {\Large \bf 
Polarization in Hadronic $\Lambda$ Hyperon Production
and Chiral-Odd Twist-3 Distribution
  \\}

\vspace{1cm}
 {\sc Y.~Kanazawa and Yuji~Koike}
\\[0.3cm]
\vspace*{0.1cm}{\it Department of Physics, Niigata University,
Ikarashi, Niigata 950--2181, Japan}
\\[1cm]


  \vskip1.8cm
  {\large\bf Abstract:\\[10pt]} \parbox[t]{\textwidth}{

Polarization of the $\Lambda$ hyperon 
produced with a large transverse momentum in the
unpolarized nucleon-nucleon collision
is analyzed in the framework of QCD factorization.  
We focus on the mechanism
in which the soft-gluon component of the 
chiral-odd spin-independent twist-3 
quark distribution $E_{F}(x,x)$ becomes a source of the polarized quark
fragmenting into the polarized $\Lambda$.  
Our simple model estimate for this contribution 
indicates that it gives rise to a significant
$\Lambda$ polarization
at large $x_F$.  This is in parallel with the observation that
the soft gluon pole mechanism gives rise to a large
single transverse spin asymmetry in the pion production at $x_F\to 1$.  
}

\end{center}

\vskip1cm

\noindent
PACS numbers: 12.38.-t, 12.38.Bx, 13.85.Ni, 13.88.+e

\noindent
[Keywords: $\Lambda$ polarization, Twist three, Chiral-odd]

  \vskip1cm 

\end{titlepage}

\setcounter{equation}{0}

It is a well known experimental fact that the hyperons produced 
in the unpolarized nucleon-nucleon collisions are polarized transversely 
to the production plane\,\cite{Bunce76,Smith87,Ho90}.
In this letter we focus on the polarization 
for the $\Lambda$ hyperon production with 
large transverse momentum ($l_T$) in $pp$ collision
\begin{eqnarray}
N(P) + N'(P') \to \Lambda(l,\vec{S}_{\perp}) + X.
\label{pplam}
\end{eqnarray}  
Ongoing experiment at RHIC is expected to provide more data
and shed light on the mechanism for the polarization.
As in the case of single transverse-spin asymmetries
in the direct photon or pion production,
$p^{\uparrow}+p\rightarrow \{\gamma,\pi \}+ X$,
nonzero $\Lambda$ polarization in the above process (\ref{pplam})
indicates the effect of 
particular quark-gluon (higher twist) correlations 
and/or the transverse momentum of partons 
either in the unpolarized nucleon or in the fragmentation function for
$\Lambda$\,\cite{ET82}-\cite{ABAM00}. 
According to the generalized QCD factorization theorem\,\cite{CSS89},
the polarized cross section for this process
consists of two types of twist-3 contributions:
\begin{eqnarray}
&{\rm (A)}&\quad E_a(x_1,x_2)\otimes q_b(x')\otimes \delta D_{c\to \Lambda}
(z)\otimes\hat{\sigma}_{ab\to c},
\label{aterm}\\
&{\rm (B)}&\quad
q_a(x)\otimes q_b(x') \otimes  D_{c\to \Lambda}^{(3)}(z_1,z_2)\otimes 
\hat{\sigma}_{ab\to c}'
\label{bterm}.
\end{eqnarray}  
Here the functions $E_a(x_1,x_2)$ and $D^{(3)}_{c\to\Lambda}(z_1,z_2)$
are the twist-3 quantities representing, respectively, the
unpolarized distribution and the fragmentation function for the transversely
polarized $\Lambda$ hyperon, 
and $a$, $b$ and $c$ stand for the parton's species.
Other functions are twist-2:  
$q_b(x)$ the unpolarized distribution (quark or gluon) and
$\delta D_{c\to \Lambda}(z)$ the transversity fragmentation function 
for the $\Lambda$. 
The symbol $\otimes$ denotes convolution. 
$\hat{\sigma}_{ab\to c}$ and
$\hat{\sigma}_{ab\to c}'$ 
represent the partonic cross section
for the process $a+b \to c + anything$ which yields large transverse 
momentum of the parton $c$. 
Note that (A) contains two chiral-odd functions
$E_a$ and $\delta D_{c\to \Lambda}$, 
while (B) contains only chiral-even functions. 

In this report, we derive a QCD formula for the polarized cross section 
(\ref{pplam}) from the (A) term which becomes dominant
in the kinematic region $|x_{F}|\to 1$,
using the valence quark-soft gluon approximation proposed by Qiu and 
Sterman\cite{QS99}.
Employing this approximation, 
they reproduced
the E704 data for the single transverse-spin 
asymmetries in the pion production at $x_F \to 1$ 
reasonably well.
The fact that the perturbative QCD description
for the pion production is valid at the pion transverse momenta as low as 
a few GeV
encouraged us to
apply the method
to the polarized $\Lambda$ hyperon production (\ref{pplam})
for which the data exist only in the same low $l_T$ region.  
At large $x_F>0$, 
where large $x$- and small $x'$- region of the parton distributions
is mainly probed,
the cross section is dominated by the particular terms in (A)
which contain the derivatives of the {\it valence} twist-3 distribution
$E_{Fa}(x,x)$.  The reason for this observation is 
the relation 
$|{\partial \over \partial x}E_{Fa}(x,x)| \gg E_{Fa}(x,x)$
owing to the behavior of $E_{Fa}(x,x)
\sim (1-x)^\beta$ ($\beta >0$) at $x\to 1$.
We thus keep only the terms with the derivative of $E_{Fa}(x,x)$ 
for the valence 
quark ({\it valence quark-soft gluon approximation}). 

The polarized cross section for (\ref{pplam}) is a function of 
three independent variables, 
$S=(P+P')^2\simeq 2P\cdot P'$, 
$x_F = 2l_{\parallel}/ \sqrt{S}$ ($=(T-U)/S$), 
and $x_T = 2l_{T}/ \sqrt{S}$.    
$T=(P-l)^2\simeq -2P\cdot l$ and 
$U=(P'-l)^2\simeq -2P'\cdot l$ are given in terms of these three
variables by 
$T= -S\left[ \sqrt{x_F^2 + x_T^2} - x_F\right]/2$ and 
$U= -S\left[ \sqrt{x_F^2 + x_T^2} + x_F\right]/2$.
In this convention, production of $\Lambda$
in the forward hemisphere in the direction of the incident nucleon
($N(P)$) corresponds to $x_F>0$.  
Since $-1<x_F <1$, $0<x_T<1$ and $\sqrt{x_F^2 + x_T^2} < 1$,
$x_F \to 1$ corresponds to the region with $-U\sim S$ and $T\sim 0$.

The cross section for 
(\ref{pplam}) from the (A) term can be derived by the method
described in \cite{QS91,QS99,KK00}. 
In the valence quark-soft gluon approximation, it 
is obtained as, 
\begin{eqnarray}
E_l{d^3\Delta\sigma^A(S_{\perp}) \over dl^3}
&=&{\pi M\alpha_s^2 \over S}\sum_{a,c}\int_{z_{min}}^1
{d\,z\over z^3}{\delta D}_{c\to\Lambda}(z)
\int_{x_{min}}^1 {d\,x\over x}
{1\over xS + U/z}
\nonumber\\
& \times & 
\int_0^1 {d\,x'\over x'}\delta\left(x'+{xT/z \over xS + U/z}\right)
\varepsilon_{l S_{\perp} p n}
\left({1\over -\hat{u}}\right)
\left[ -x {\partial \over \partial x}E_{Fa}(x,x)\right]
\nonumber\\
& \times &
\left[ 
G(x')\delta \widehat{\sigma}_{ag\to c} +\sum_{b} q_b(x') 
\delta \widehat{\sigma}_{ab\to c}
\right]\delta_{ac},
\label{cross}
\end{eqnarray}
where 
$p$ and $n$ are the two light-like vectors defined from the momentum of the
unpolarized nucleon (mass $M$) as $P=p+M^2n/2$, $p\cdot n=1$ and
$\varepsilon_{l S_{\perp} p n}=\varepsilon_{\mu\nu\lambda\sigma}l^\mu 
S_{\perp}^{\nu} p^\lambda n^\sigma\sim {\rm sin}\phi$ with $\phi$ 
the azimuthal angle between the spin vector of the $\Lambda$ hyperon and 
the production plane.  
The invariants in the parton level are defined as
\begin{eqnarray}
\hat{s} &=&(p_a + p_b)^2 \simeq ( xP + x'P')^2 \simeq xx'S,\nonumber\\
\hat{t} &=&(p_a-p_c)^2 \simeq (xP-l/z)^2 \simeq xT/z,\nonumber\\
\hat{u} &=&(p_b-p_c)^2 \simeq (x'P'-l/z)^2 \simeq x'U/z.   
\end{eqnarray}
The lower limits for the integration variables are 
\begin{eqnarray}
z_{min} = {-(T+U) \over S}=\sqrt{x_F^2+x_T^2},\qquad
x_{min} = {-U/z \over S+T/z}.  
\end{eqnarray}
$q_b(x')$ and $G(x')$ are the unpolarized quark and gluon distributions, 
respectively. 
$\delta \widehat{\sigma}_{ag\to c}$ and $\delta \widehat{\sigma}_{ab\to c}$ 
are the partonic cross sections 
for the quark-gluon and quark-quark subprocesses,  
respectively. 
$E_F(x,x)$ is the soft gluon component of the
unpolarized twist-3 distribution defined as
\begin{eqnarray}
E_{Fa}(x,y)={-i\over 2M}
\int{d\lambda \over 2\pi}e^{i\lambda x}\langle P | \bar{\psi}^a(0)\!\not{\! n}
\gamma_{\perp\sigma}
\left\{ \int {d\mu\over 2\pi}e^{i\mu (y-x)}
g F^{\sigma\beta}(\mu n)n_\beta\right\}
\psi^a(\lambda n)|P\rangle.  
\label{EF}
\end{eqnarray}
$\delta D_{c\to\Lambda}(z)$ is the twist-2 transversity fragmentation 
function for $\Lambda$ defined by
\begin{eqnarray}
\delta D_{c\to\Lambda}(z)=
\sum_X {z\over 4}\int{d\lambda\over 2\pi}e^{-i\lambda/z}
\langle 0|\gamma_5 \!\not{\! S}_\perp \!\not{\! n}_\Lambda
\psi^c(0)|\Lambda(lS_\perp)X\rangle \langle \Lambda(lS_\perp) 
X|\bar{\psi}^c(\lambda 
n_\Lambda)|0\rangle, 
\end{eqnarray}
where the lightlike vector $n_\Lambda$ is an analogue of $n$ for $l$
($l\cdot n_\Lambda =1$). 
The summation for the flavor indices of $E_{Fa}(x,x)$ is to be over 
$u$- and $d$- valence quarks,   
while that for the twist-2 distributions is 
over $u$, $d$, $\bar{u}$, $\bar{d}$,
$s$, $\bar{s}$.
The missing contributions in (\ref{cross}) are
the soft gluon pole contribution proportional to
$E_{Fa}(x,x)$ itself (without derivative) and the soft Fermion pole 
contribution proportional to $E_D(x,0)$ which does not appear
with the derivative.   $E_D(x,y)$ is
obtained by the replacement of the gluon field strength $gF^{\mu\nu}n_\nu$
by the covariant derivative $D^\mu$ in $E_F(x,y)$ (see \cite{KK00}).
As stated above, at large $x_F$ (i.e. large $x$) 
$(d/dx)E_F(x,x)$ receives an enhancement, and (\ref{cross}) becomes the
most dominant contribution.  At large $x_F$, $z$ is also large and 
we anticipate that 
the term proportional to $(d/dz)D_{c\to\Lambda}^{(3)}(z,z)$
in the (B) contribution (\ref{bterm}) also brings another
large contribution.
We left the analysis of this term for future study.

After the soft gluon poles are properly handled, 
$\delta\widehat{\sigma}_{ab\to c}$ and  
$\delta\widehat{\sigma}_{ag\to c}$ in (\ref{cross}) 
can be obtained from the $2\to 2$ cut diagrams
shown in Figs. 1 and 2, respectively\,\cite{QS91,QS99,KK00}.
In these figures, the quark lines labeled as $a$ comes from 
$E_{Fa}(x,x)$ and the quark line labeled as $c$ fragments into
$\Lambda$.  
Because of the chiral-odd nature of $E_{Fa}$ and $\delta D_{c\to \Lambda}$
they have to appear in a pair along a Fermion line, and hence 
only four diagrams in Fig. 1 
contribute to $\delta\widehat{\sigma}_{ab\to c}$.  
The result for the hard cross section reads 
\begin{eqnarray}
\delta\widehat{\sigma}_{q q'\to q}
&=&
\left({\widehat{s}\hat{u}\over \hat{t}^2}\right)
\left[{2\over 9}+{1\over 9}\left(1+{\hat{u}\over \hat{t}}\right)\right],
\qquad({\rm Fig.\ 1(a)})
\nonumber\\
\delta\widehat{\sigma}_{q \bar{q}'\to q}
&=&
\left({\hat{s}\hat{u}\over \hat{t}^2}\right)
\left[{7\over 9}+{1\over 9}\left(1+{\hat{u}\over \hat{t}}\right)\right],
\qquad({\rm Fig.\ 1(b)})
\nonumber\\
\delta\widehat{\sigma}_{q q\to q}
&=&
-\left({\hat{s}\over \hat{t}}\right)
\left[{10\over 27}+{1\over 27}\left(1+{\hat{u}\over \hat{t}}\right)\right],
\qquad({\rm Fig.\ 1(c),(d)})
\label{qqhard}
\end{eqnarray}
for $\delta\widehat{\sigma}_{ab\to c}$, and 
\begin{eqnarray}
\delta\widehat{\sigma}_{q g\to q}
&=&
{9\over 8}\left({\hat{s}\hat{u} \over \hat{t}^2}\right)
+{9\over 8}\left({\hat{u} \over \hat{t}}\right)+{1\over 8}
+\left[{1\over 4}\left({\hat{s}\hat{u}\over \hat{t}^2}\right)
+{1\over 72}\right]
\left(1+{\hat{u}\over\hat{t}}\right),
\label{qghard}
\end{eqnarray}
for $\delta \widehat{\sigma}_{ag\to c}$.  
We note that in the large $x_F$ region $\hat{t}$ becomes small
compared to $\hat{s}$ and $\hat{u}$.  Therefore
$\delta\widehat{\sigma}_{ab\to c}$ (from Figs. 1 (a),(b)) and  
$\delta\widehat{\sigma}_{ag\to c}$ (from Fig. 2) contributes to a large
polarization as was the case for the single transverse-spin
asymmetry for the pion production\,\cite{QS99}. 
This is in strong contrast to the chiral-odd contribution for the 
single transverse-spin asymmetry for the pion production studied in 
\cite{KK00}:  In the contribution identified in \cite{KK00},
two chiral-odd distributions come 
from the initial nucleons and have to form a closed quark loop in a pair.
Therefore there is no contribution
from Figs. 1(a),(b) and Fig. 2, leading to the negligible asymmetry. 
(See the second reference of \cite{KK00}.)

We now present a simple estimate of the $\Lambda$ polarization $P_\Lambda$. 
To this end we employ a model for $E_F(x,x)$ introduced in Ref. \cite{KK00}. 
It is based on 
the comparison of the explicit form (\ref{EF})
with the transversity distribution
\begin{eqnarray}
\delta q_a(x)&=& {i\over 2}\varepsilon_{S_\perp \sigma p n}
\int{d\lambda \over 2\pi}e^{i\lambda x}\langle PS |\bar{\psi}^a(0)\!\not{\! n}
\gamma_\perp^\sigma \psi^a(\lambda n)|PS\rangle,
\label{transversity}
\end{eqnarray}
where $\varepsilon_{S_\perp \sigma p n}\equiv \varepsilon_{\mu\sigma\nu\lambda}
S_\perp^\mu p^\nu n^\lambda$.
We make an ansatz 
\begin{eqnarray}
E_{Fa}(x,x) = K_a \delta q_a(x),
\label{EFq}
\end{eqnarray}
with a flavor-dependent parameter $K_a$ which simulates
the effect of the gluon field with zero momentum in $E_F(x,x)$.
We note that even though $E_F(x,x)$ is an unpolarized distribution,
the quarks in $E_F(x,x)$ is ``transversely polarized'' which
eventually fragments into the transversely polarized $\Lambda$. 
The relation (\ref{EFq}) is in parallel with the ansatz originally introduced
in \cite{QS99},
\begin{eqnarray}
G_{Fa}(x,x) = K_a' q_a(x),
\label{QSassumption}
\end{eqnarray}
which is also motivated by the explicit forms for
$G_{Fa}(x,x)$ and $q_a(x)$:
\begin{eqnarray}
G_{Fa}(x,x)={1\over M}\varepsilon_{S_\perp \sigma p n}
\int{d\lambda \over 2\pi}e^{i\lambda x}\langle P | \bar{\psi}^a(0)\!\not{\! n}
\left\{ \int {d\mu\over 2\pi}g F^{\sigma\beta}(\mu n)n_\beta\right\}
\psi^a(\lambda n)|P\rangle, 
\label{GF}
\end{eqnarray}
\begin{eqnarray}
q_a(x)&=& {1\over 2}
\int{d\lambda \over 2\pi}e^{i\lambda x}\langle PS |\bar{\psi}^a(0)\!\not{\! n}
\psi^a(\lambda n)|PS\rangle.
\end{eqnarray}
From these relations,
we expect that the signs of $K_a$ and $K'_a$ are the same and
their magnitude is similar.
Using the unpolarized parton distribution in \cite{GRV95} (scale
$\mu=1.5$ GeV)
and the fragmentation function for the pion in \cite{BKK95}
(scale $\mu=2$ GeV), 
$K'_{u,d}$ was determined to be $K'_{u}=-K'_{d}=0.06$ so that it
approximately reproduce
the FNAL E704 data of the single transverse-spin 
asymmetry $A_N$ in the pion production\,\cite{Adams91} at large $x_F$.

In calculating the $\Lambda$ polarization, 
we use each distribution and fragmentation function
at the scale $1.1$ GeV which is equal to the
average
transverse momentum $l_T$ of the produced $\Lambda$
in CERN R608 data\,\cite{Smith87}. 
For the unpolarized distribution $q_a(x)$ and $G(x)$,
we use the GRV LO distribution\,\cite{GRV95}
(same distribution to determine $K'_{u,d}$ from $A_N$).
For the transversity distribution $\delta q_a(x)$,
we use the GRSV helicity distribution $\Delta q_a(x)$
(LO, standard scenario)\cite{GRSV96} assuming 
$\delta q_a(x)=\Delta q_a(x)$ at the scale $\mu=1.1$ GeV. 
Fragmentation functions of $\Lambda$ are taken from 
Ref.\,\cite{FSV98}: 
For the transversity fragmentation function $\delta D_{c\to\Lambda}(z)$
we use longitudinally polarized fragmentation
$\Delta D_{c\to\Lambda}(z)$ (three different scenarios) assuming
$\delta D_{c\to\Lambda}(z)
=\Delta D_{c\to\Lambda}(z)$ at the scale $\mu=1.1$ GeV.
We note that the $\Lambda$ fragmentation functions
given in \cite{FSV98} is for $\Lambda+\overline{\Lambda}$. 
But in the kinematic region of our interest
(large $x_F$ $\sim$ large $z$),  
$D_{u,d\to\Lambda +\overline{\Lambda}}(z)$
can be regarded as the one for $\Lambda$ itself,
so that we use $D^\Lambda$ in \cite{FSV98}
for the $\Lambda$ fragmentation function. 
With these preparations,
our first choice for $K_{u,d}$ is to set
$K_{u,d}=K_{u,d}'$ as noted above.  
We also show the result for $K_u=-K_d=0.24$ which is
determined to approximately reproduce the 
R608 data on $P_\Lambda$ (see below).

The calculated $\Lambda$ polarization $P_\Lambda$ 
obtained with $K_{u,d}=K'_{u,d}$  
is shown in Fig. 3 for the three scenarios of 
$\delta D_{c\to\Lambda}(z)$ in \cite{FSV98}
together with the CERN R608 data.
The scenario 1 corresponds to the expectation from the naive 
non-relativistic 
quark model, where only strange quarks can fragment into a 
polarized $\Lambda$. 
In our approximation, $E_F(x,x)=0$ for the $s$-quark and thus 
the polarization is zero in this scenario. 
The scenario 2 is based on the assumption that the flavor-dependence of 
$\Delta D_{c\to\Lambda}(z)$ is the same as that of  
the polarized quark distribution in $\Lambda$
obtained by the SU(3) symmetry from $g_1^p$ data;
$\delta D_{u\to \Lambda}=
\delta D_{d\to \Lambda}=-0.2\delta D_{s\to \Lambda}$.  
In the scenario 3, three flavors of quarks equally fragment into
the polarized $\Lambda$; 
$\delta D_{u\to\Lambda}=\delta D_{d\to\Lambda}=\delta D_{s\to\Lambda}$. 
From Fig. 3, one sees that
the scenarios 2 and 3 give rise to increasing polarization at large $x_F$
as expected, the former giving the same sign of $P_\Lambda$ as
the data.   To get negative $P_\Lambda$, negative 
$\delta D_{u,d\to\Lambda}(z)$
is necessary, 
if the (A) term turns out to be responsible for the
hyperon polarization.  
To get an 
approximate fit to the data with the scenario 2
for $\delta D(z)$, one needs $K_u=-K_d=0.24$.  This curve is also shown 
by the solid line 
in Fig. 3.  At intermediate $x_F$, we expect a moderate
contribution from the nonderivative term of the soft-gluon pole
as well as the soft-fermion pole contributions, which may
bring better
agreement with the data.
	
The curves in Fig. 3 are obtained by the ansatz
for the transversity distribution and fragmentation functions, 
$\delta q(x) = \Delta q(x)$,
$\delta D(x) = \Delta D(x)$.  
If we had used saturation assumption in the inequality \cite{Soffer}, 
$2|\delta q(x)| \leq q(x) + \Delta q(x)$,
$2|\delta D(x)| \leq D(x) + \Delta D(x)$,
we would have got smaller $K_{u,d}$ than estimated above.  

Since we kept only valence distribution for $E_F(x,x)$, 
polarization of $\overline{\Lambda}$ becomes identically zero in all
three scenarios which is consistent with the experimental data.
(For other anti-hyperons such as 
$\overline{\Sigma}$ and $\overline{\Xi}$, however, 
nonzero polarization 
has been experimentally observed\,\cite{Ho90}, so that the 
non-valence component of the distribution function is certainly necessary for
the complete description of the hyperon
polarization.)

In Fig. 4, we plotted the $\Lambda$ polarization
for $\sqrt{S}=20,\ 40,\ 200$ GeV at $l_{T}=1.1$ GeV with the scenario 2
for $\delta D_{c\to \Lambda}$ and $K_u=-K_d=0.24$.  
One sees that the 
result is almost independent of $\sqrt{S}$.
This tendency is the same as the experimental data.

The hyperon polarization shown above shares common
features with the single transverse spin-asymmetry $A_N$ in the pion
production, $p^\uparrow + p \rightarrow \pi(l_T) + X$
studied in \cite{QS99}. 
With our model assumption (\ref{EFq}) for $E_F(x,x)$, 
the approximate formula for the $\Lambda$ polarization at large $x_F$ 
can be written as
\begin{eqnarray}
P_\Lambda \sim
{K\pi M l_T \over (-U)}\left[ 1 + O\left( {U\over T}\right)\right]
{1\over 1-x_F}{\delta q(x) \otimes \delta D(z) \over 
 q(x) \otimes D(z)}.
\label{approx}
\end{eqnarray}
Here the factor $l_T/(-U)$ comes from
$\varepsilon_{lS_\perp p n}/(-\hat{u})$ in
(\ref{cross}), $O(U/T)$ term is from those with
$\hat{u}/\hat{t}$ in (\ref{qqhard}) and (\ref{qghard})
with the coefficients smaller than for the other terms and
$1/(1-x_F)$ dependence comes from the derivative of $E_F(x,x)$.
The last factor represents the ratio of 
parton distribution and fragmentation functions
between polarized and unpolarized production cross section, 
which is absent in the analogous formula for $A_N$\,\cite{QS99}.  
$P_\Lambda$ obtained
with the assumption $K_{u,d}=K'_{u,d}$ in the scenario 2 is 
smaller than $A_N$, which is mainly due to the small ratio
$\delta D(z)/D(z)$.  
At $x_F >> x_T$, $U\to -x_FS$ and $T\to -l_T^2/x_F$.
Accordingly,
$P_\Lambda$ has typically two contributions proportional to
$l_T/S$ and $1/l_T$ corresponding to
$O(l_T/(-U))$ and $O(l_T/(-T))$ terms in (\ref{approx}),
respectively. 
Fig. 5 shows the $l_T$-dependence of the $\Lambda$ polarization
at $x_F=0.7$ for $\sqrt{S}=20,\ 40,\ 200$ GeV.  
The polarization changes rapidly 
at $l_T< 2$ GeV, which indicates the second term in (\ref{approx})
becomes dominant in this region.
Experimentally, 
$P_\Lambda$ tends to decrease as $l_T$ increases 
in the region $l_T \sim 1-3$ GeV at $x_F > 0.6$, 
while it shows flat $l_T$ dependence in the smaller $x_F$ region. 
At $\sqrt{S}=20$ GeV, $P_\Lambda$ increases slightly as $l_T$ increases
toward the edge of the phase space ($\sim$ 7 GeV), as was seen in 
$A_N$ \cite{QS99}.

We remark that the effect of the $\Lambda$ mass $M_\Lambda$ 
is completely ignored in our formula for $P_\Lambda$. 
The rapid increase of $P_\Lambda$ at small $l_T$
is due to our unjustified neglect of $M_\Lambda$ compared with $l_T$. 
In particular, $P_\Lambda$ should be
zero at $l_T =0$.
The comparison of our formula with experimental data should be more ideally
done at larger $l_T$ as well as large $x_F$ in future collider experiments.

The complete formula for the hyperon polarization receives
the (B) contribution in (\ref{bterm}).
Analysis of this term involves some complications
due to the presence of two kinds of hadron matrix elements
for the twist-3 fragmentation function.
We hope to present the analysis of this term in a separate publication. 
At this stage we can only say that the (A) contribution in (\ref{aterm})
brings a significant polarization of $\Lambda$ at large $x_F$,
if the soft-gluon pole mechanism analyzed in \cite{QS99}
is a major source for the single transverse-spin asymmetry
in the pion production.

A different approach to the hyperon polarization
introduces
the so-called T-odd distribution or 
fragmentation functions with the intrinsic
transverse momentum\,\cite{Sivers,Collins,ABAM00,MT96} instead of 
twist-3 distributions 
introduced here.
Similarly to (A) and (B),
this approach starts from the factorization assumption for the two
types of contributions to the polarization;
(i) $h_{1}^\perp (x,{\bf p}_\perp)\otimes q(x')\otimes \delta D(z)\otimes
\hat{\sigma}$,
(ii) $q(x)\otimes q(x')\otimes D_{1T}^{\perp} (z,{\bf k}_{\perp})\otimes
\hat{\sigma}'$,
where 
$h_1^\perp$ represents distribution of a transversely polarized quark 
with nonzero tranverse mometum inside the unpolarized nucleon, 
and $D_{1T}^{\perp}$ represents a fragmentation function for 
an unpolarized quark fragmenting into a transversely polarized $\Lambda$ 
with the transverse momentum (``polarizing fragmentation function''). 
Anselmino {\it et al.} fitted the experimental data for the $\Lambda$ 
polarization assuming
the above (ii) is the sole origin of the polarization\cite{ABAM00}.
Experimentally, however, no significant transverse polarzation of $\Lambda$
has been observed in $e^+ + e^-\rightarrow \Lambda^\uparrow +X$\,\cite{Aleph},
in which origin of the $\Lambda$ polarization
resides only in the fragmentation process.  
This data suggests that  
the whole $\Lambda$ polarization may not be ascribed to the 
$D_{1T}^\perp$-effect in the hadronic production.
\footnote{A recent paper\,\cite{Boer01} ascribed the smallness
of the $\Lambda$ polarization in \cite{Aleph} to the Sudakov
suppression.}    
We also expect from the present study that the
significant portion of the $\Lambda$ polarization should be ascribed 
to the twist-3 distribution in the unpolarized nucleon
and $\delta D(z)$ which should be related to 
the above contribution (i).
It is also interesting to explore the connection between 
the present approach and that in \cite{ABAM00}.  

To summarize, 
we have derived a cross section formula for the polarized $\Lambda$ 
production in the unpolarized nucleon-nucleon collision at large $x_{F}$.  
We focused on the mechanism where the 
soft-gluon component of the unpolarized twist-3 quark distribution 
becomes the source of the polarized quark fragmenting into 
the polarized $\Lambda$. 
A simple model estimate for this contribution suggests
a possibility that the contribution from the soft gluon pole 
gives significant $\Lambda$ polarization.  

\vspace{1cm}

{\noindent\Large\bf Acknowledgement}
\vspace{1ex}

We thank D. Boer, J. Qiu and G. Sterman for useful discussions and
W. Vogelsang for providing us with their Fortran code for the
$\Lambda$ fragmentation functions.  This work is supported in part
by the Grant-in-Aid for Scientific Research (No. 12640260) of the Ministry
of Education, Science, Sports and Culture (Japan).

\newpage

\large
\centerline{\bf Figure Captions}

\normalsize
\begin{enumerate}

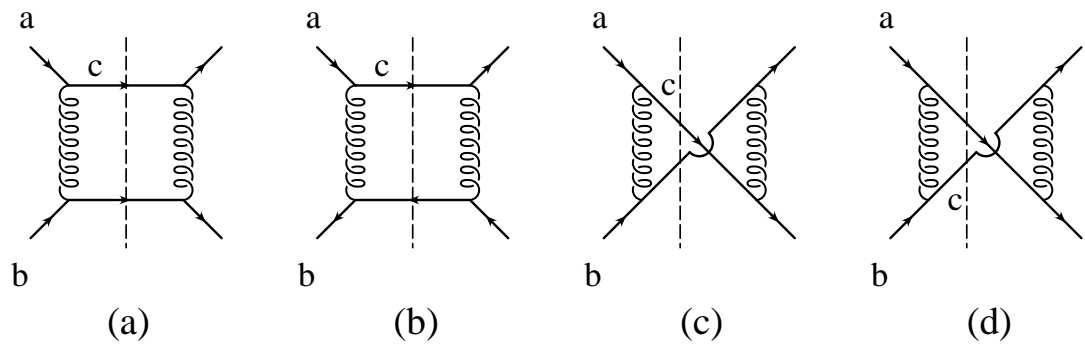
\item[{\bf Fig. 1}]
Quark-quark $2\to 2$ scattering diagrams contributing to the 
hard cross section.

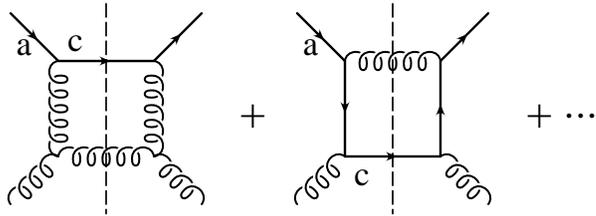
\item[{\bf Fig. 2}]
Quark-gluon $2\to 2$ scattering diagrams contributing to the 
hard cross section.

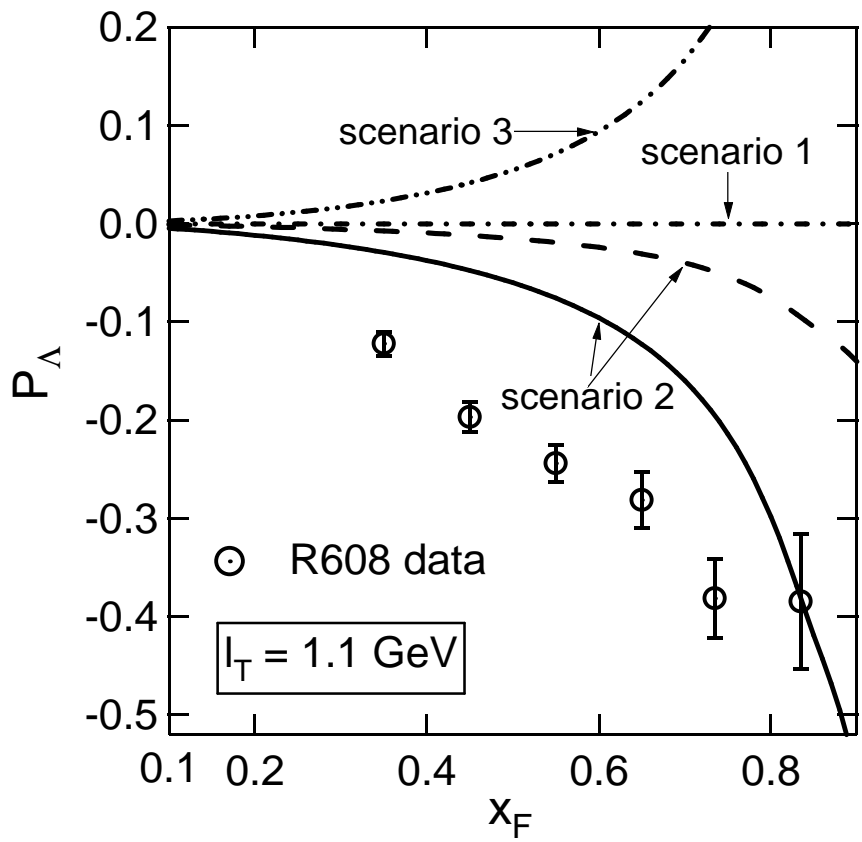
\item[{\bf Fig. 3}]
The calculated 
$\Lambda$ polarization $P_{\Lambda}$
with three scenarios for $\delta D_{c\to\Lambda}$
at $\sqrt{S}=62$ GeV and the transverse momentum of $\Lambda$ at $l_T
=1.1$ GeV together with the R608 data\,\cite{Smith87}.  
For the scenario 2, the result 
with $K_{u,d}=K'_{u,d}$ 
and $K_u=-K_d =0.24$ are shown by dashed and solid lines,
respectively.
For the scenarios 1 and 3, only the result with
$K_{u,d}=K'_{u,d}$ is shown by dash-dotted and dash-double-dotted 
lines, respectively.

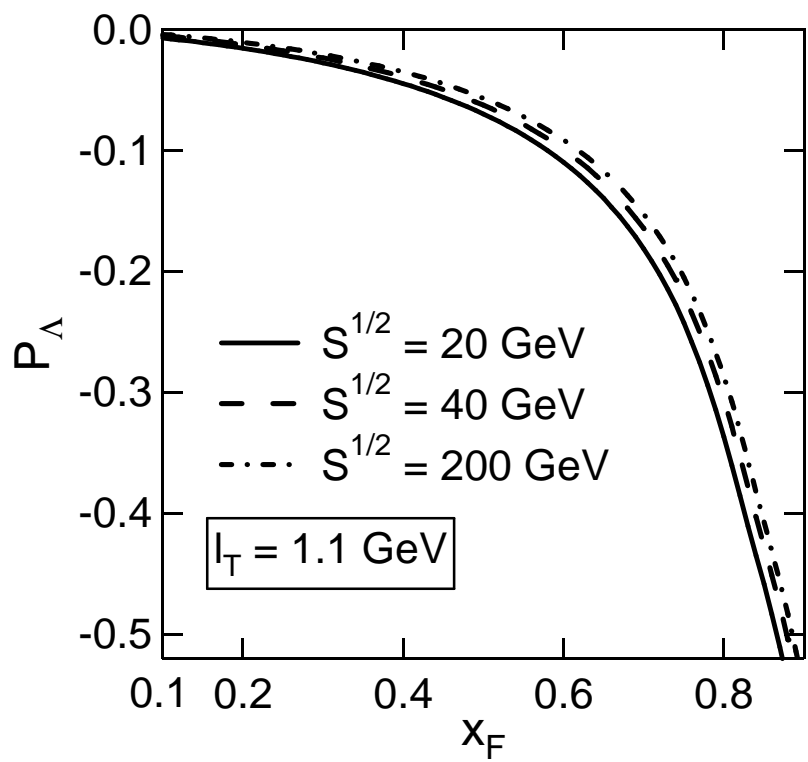
\item[{\bf Fig. 4}]
The $\Lambda$ polarization $P_{\Lambda}$ 
at $\sqrt{S}=20,\, 40,\,200$ GeV and $l_T=1.1$ GeV. 
Scenario 2 is used for $\delta D_{c\to\Lambda}$ and
$K_u=-K_d=0.24$.

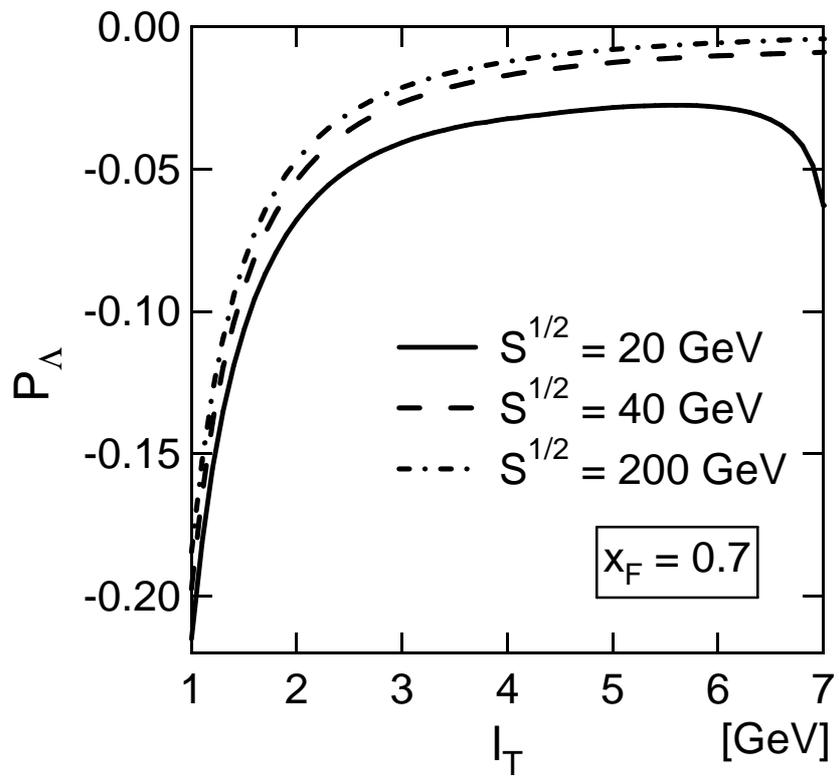
\item[{\bf Fig. 5}]
$l_T$-dependence of 
$P_{\Lambda}$ 
at $\sqrt{S}=20,\,40,\,200$ GeV and $x_{F}=0.7$.
Scenario 2 is used for $\delta D_{c\to\Lambda}$ and
$K_u=-K_d=0.24$.

\end{enumerate}

\end{document}